\documentclass[sigconf]{acmart}
\AtBeginDocument{%
  }

\setcopyright{rightsretained}
\copyrightyear{2026}
\acmYear{2026}
\acmConference[ICSE-SEET]{}{}

\begin{document}

\title{Amplifiers or Equalizers? A Longitudinal Study of LLM Evolution in Software Engineering Project-Based Learning}

\author{Hana Kataoka}
\affiliation{%
  \institution{Nihon University}
  \city{Tokyo}
  \country{Japan}}
\email{cshn24004@g.nihon-u.ac.jp}

\author{Jialong Li}
\authornote{Corresponding Author: Jialong Li}
\affiliation{%
  \institution{Waseda University}
  \state{Tokyo}
  \country{Japan}}
\email{lijialong@fuji.waseda.jp}

\author{Yutaka Matsuno}
\affiliation{%
  \institution{Nihon University}
  \city{Tokyo}
  \country{Japan}}
\email{matsuno.yutaka@nihon-u.ac.jp}

\begin{abstract}
As LLMs reshape software development, integrating LLM-augmented practices into SE education has become imperative. While existing studies explore LLMs' educational use in introductory programming or isolated SE tasks, their impact in more open-ended Project-Based Learning (PBL) remains unexplored. This paper introduces a two-year longitudinal study comparing a 2024 (using early free LLMs, $n$=48) and 2025 (using the latest paid LLMs, $n$=46) cohort. Our findings suggest the latest powerful LLMs' dual role: they act as "equalizers," boosting average performance even for programming-weak students, providing opportunities for more authentic SE practices; yet also as "amplifiers," dramatically widening absolute performance gaps, creating new pedagogical challenges for addressing educational inequities.
\end{abstract}


\maketitle

\section{Introduction}
Project-based learning (PBL) is a widely adopted approach in software engineering (SE) education \cite{10.1109/ICSE-SEET58685.2023.00015, 10.1145/3639474.3640074, 9768199, 10.1109/TE.2021.3137344}. By engaging students in the creation of software artifacts following SE methodology, PBL provides an authentic context for them to apply theoretical knowledge and experience the entire development lifecycle, from requirements definition to testing. Inquiry PBL (IPBL), where students define their own problems and solutions, is particularly valued because it can foster greater intrinsic motivation and ownership compared to projects with predetermined specifications~\cite{Singh_2020,Ahmed2025Inquiry}.

Concurrently, the landscape of software development is being reshaped by the remarkable capabilities of Large Language Models (LLMs)~\cite{10.1145/3695988,10.1145/3686803}. Given their proficiency in code generation and other SE activities, LLMs have been rapidly adopted across the software industry, and the ability to leverage LLM tools is fast becoming an essential competency for software engineers~\cite{10.1145/3660788, HAQUE2025100204, Pauklin2025AIGenCode, Nanthakumar2025AIRevolution}. This industry shift places a clear mandate on SE education to evolve and incorporate LLM-augmented activities, not merely as an add-on, but as an integral part of preparing students for their future careers~\cite{khan2025integratingllmssoftwareengineering,sengul2024software,10.1145/3626252.3630927}.

However, while the educational use of LLMs in introductory programming~\cite{10.1145/3639474.3640059, 10.1145/3639474.3640076} or specific SE tasks~\cite{10.1145/3639474.3640058,10.1145/3639474.3640052,11024401,11024458,10662984,10662994,11024341,khan2025integratinglargelanguagemodels} has been explored, their impact within the context of SE PBL remains largely unknown. Unlike well-defined programming exercises or isolated tasks, PBL grants students significant freedom and demands more sophisticated and adaptive use of LLMs.
This inherent complexity increases the difficulty of pedagogical design when integrating LLMs in PBL, especially given the ongoing advancement of LLM evolution: 
How has the rapid LLM evolution reshaped student performance in PBL? What new opportunities do the latest LLMs unlock, and what novel challenges do they introduce?

To this end, this paper introduces a two-year longitudinal empirical study, rooted in a five-week IPBL assignment in a third-year undergraduate SE course. Specifically, we compare a 2024 cohort where students mainly used free, publicly available LLMs, and a 2025 cohort that used more powerful, paid LLMs. By analyzing the performance across these two groups, we aim to provide the initial empirical evidence on the multifaceted impacts of the rapid evolution of LLM in SE PBL.

\section{Study Design and Context}

\textit{Course Context}.
The study was conducted based on a mandatory 15-week SE course for third-year undergraduate computer science students at a local university.
All participants completed foundational programming, data structures, and algorithms courses, providing necessary technical skills for development.
The course consists of ten weeks of lectures on software development lifecycle fundamentals, followed by five weeks of individual IPBL, where students develop complete applications of their choice.
Students are tasked to follow the waterfall model \cite{Sherrell2013} and retain all intermediate artifacts.
Course content and instructors remained unchanged between 2024 and 2025, but students' LLM usage changed: 2024 students mostly used free ChatGPT 4o, while 2025 students primarily used the more capable Gemini 2.5 Pro (Pro plan for students).

\textit{Participants}.
We recruited two student groups. The first group is the ``2024 Cohort'' (Baseline Group, $n = 48$), which completed the course from April to August 2024. The second group is the ``2025 Cohort'' (Experimental Group, $n = 46$), who completed the course from April to August 2025. 

\textit{Data Collection and Measurement}.
We assess (i) COSMIC Function Points (CFP), which capture software functional size and are measured using the official COSMIC protocol by the course instructor, who has over 20 years of teaching experience in SE~\cite{ABUALKISHIK2018179,10.1145/3143434.3143446}. Given the IPBL's purpose to develop an application prototype, we use this metric as the primary indicator for measuring student performance; 
(ii)  Pre- and Post-CFRP (Common Framework of Reference for Programming Skills), which measures students' core programming abilities to six level~\cite{10.1145/3304221.3319755};
(iii) Test Case Pass Rate: which is measured using 24 (8 normal, 8 boundary, and 8 exceptional) unit test cases for each project, and these test cases are generated using LLMs with a structured prompt.

\section{Key Results \&  Findings}
\vspace{-2.5mm}
\begin{figure}[h!t]
\centering
\includegraphics[width=1\linewidth]{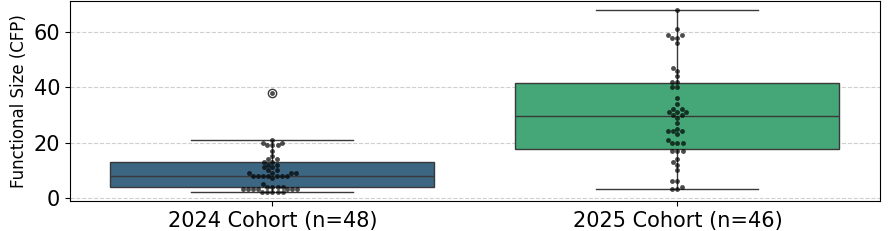}
\vspace{-7mm}
\caption{Functional Size Distribution Comparison.}
\label{fig:functional_size_comparison}
\vspace{-3mm}
\end{figure}

The most significant difference between the 2024 and 2025 cohorts was a massive increase in the functional size of the student-developed applications, as shown in Figure~\ref{fig:functional_size_comparison}.
Specifically, mean CFP surged from 9.56 to 30.13 (3.2x), while median CFP showed a pronounced jump from 8.00 to 29.50 (3.7x). 
At the same time, the relative dispersion, i.e., the Coefficient of Variation, decreased from 0.74 to 0.57.
These results provide clear, quantitative evidence that the shift to more powerful LLMs enables students to achieve significantly more ambitious and functionally rich projects than was previously possible within the same five-week timeframe, while also making their performance more homogeneous in relative terms.
However, at the same time, we also observed a dramatic expansion in the absolute performance variance. The standard deviation increased from 7.06 to 17.16 (2.4x), and the range between the maximum and minimum CFP grew substantially (from 36 to 65).

Programming proficiency (CFRP) shows a significant moderate positive correlation with reliability as measured by test pass rate (Pearson $r = 0.39$, Spearman $\rho = 0.33$, both $p < 0.01$). This suggests that fundamental skills remain crucial for reliable software, as stronger students better understand, debug, and integrate LLM-generated code. In contrast, programming proficiency has only a weak and non-significant correlation with functional size (CFP; $r = 0.16$, $\rho = 0.16$), indicating that, with the advanced LLM assistance, the ability to build large, complex applications is no longer primarily constrained by traditional programming skill levels.

\section{Discussion \& Implications}
\label{sec:discussion}

\textit{Opportunities: LLMs as Equalizers}.
As suggested by the dramatic increase in functional size and its decoupling from programming ability, we believe that superior LLMs act as an equalizer to some extent, providing an opportunity by giving most students, especially those who are not good at programming, the leverage to reach a level of project complexity where authentic SE challenges emerge. Traditionally, a five-week IPBL meant that students spent the majority of their effort on coding basic functionalities, often leaving little time for deeper application of advanced SE practices. While the primary goal of such IPBL is to familiarize students with SE concepts and development phases, the practical constraint of manual implementation often limited the scope to a point where advanced SE practices (e.g., architectural design) felt abstract or unnecessary for their ``toy'' projects.
Advanced LLMs fundamentally change this equation. By automating much of the routine code generation, they enable learning-based projects to move beyond mere conceptual familiarity toward more authentic, complex engineering practice within limited timeframes. For example, modular design becomes tangible with larger codebases; rapid prototyping enables authentic agile development practice.
In this sense, superior LLMs bring us the opportunity to provide a more realistic and advanced SE practice within the same time constraints in PBL.

\textit{Challenges: LLM as Amplifier}.
While democratizing project complexity, the move to superior LLMs simultaneously acted as a more potent amplifier of inequality, as shown in the explosion in the absolute performance gap. This means that without intervention, LLMs will further strengthen the advantages of high-performing students, while those who start behind may be further marginalized, potentially experiencing frustration from their inability to effectively harness LLMs.
Although a definitive causal analysis requires the collection of more detailed information in future work, our teaching experience suggests it may be driven by a shift in the critical skills needed in the LLM-augmented SE era. This may include systems thinking and problem decomposition abilities; prompt engineering and iterative refinement capabilities when using LLMs; and critical evaluation and integration abilities when collaborating with LLMs~\cite{etsenake2024understandinghumanllmdynamicliterature, Federiakin2024Prompt, Lee2025,FloresRomero2025,LISSACK2024389}.
In this sense, the introduction of superior LLMs brings two important challenges: (i) abilities for using LLMs, such as strategic problem decomposition and iterative prompt optimization, have become a ``hidden curriculum," creating new educational inequity; (ii) the focus of SE education must evolve from teaching traditional SE knowledge in isolation to teaching LLM-augmented SE practices, such as effective human-LLM collaboration patterns~\cite{10.1145/3706599.3719852,10.1145/3613905.3651042,10.1145/3677081}.

\section{Conclusion and Future Work}
This study introduces our two-year longitudinal empirical study on the impact of LLM evolution in IPBL. We found that powerful LLMs act as powerful equalizers by enabling students, regardless of prior proficiency, to develop more complex applications and gain a more authentic SE experience. Meanwhile, LLMs also serve as amplifiers that widen the absolute performance gap, as stronger students are enabled to produce larger and higher-quality systems. We hope our findings can inspire the redesign of LLM-era PBL and better cultivate the essential skills for the LLM-augmented SE era.

To address the threats to validity, our future work will extend this study in three directions:
(i) Conducting a more in-depth analysis of student artifacts, extending beyond functional size to assess code quality, architectural complexity, and documentation quality.
(ii) Process mining on human-LLM interaction logs to identify behavioral patterns that distinguish high-performing from low-performing students, thereby informing future pedagogical design.
(iii) Collect more comprehensive student data in the 2026 course, such as their prompt engineering proficiency, to better understand the causal factors behind the observed performance gap.

\clearpage
\bibliographystyle{ACM-Reference-Format}
\bibliography{bib}

\end{document}